# STEAM++ AN EXTENSIBLE END-TO-END FRAMEWORK FOR DEVELOPING IoT DATA PROCESSING APPLICATIONS IN THE FOG

Márcio Miguel Gomes[1], Rodrigo da Rosa Righi[1],
Cristiano André da Costa[1] and Dalvan Griebler[2]

[1]University of Vale do Rio dos Sinos - UNISINOS, RS, Brazil
[2]Pontifical Catholic University of Rio Grande do Sul, RS, Brazil

*ABSTRACT*

*IoT applications usually rely on cloud computing services to perform data analysis such as filtering, aggregation, classification, pattern detection, and prediction. When applied to specific domains, the IoT needs to deal with unique constraints. Besides the hostile environment such as vibration and electric-magnetic interference, resulting in malfunction, noise, and data loss, industrial plants often have Internet access restricted or unavailable, forcing us to design stand-alone fog and edge computing solutions. In this context, we present STEAM++, a lightweight and extensible framework for real-time data stream processing and decision-making in the network edge, targeting hardware-limited devices, besides proposing a micro-benchmark methodology for assessing embedded IoT applications. In real-case experiments in a semiconductor industry, we processed an entire data flow, from values sensing, processing and analysing data, detecting relevant events, and finally, publishing results to a dashboard. On average, the application consumed less than 500kb RAM and 1.0% of CPU usage, processing up to 239 data packets per second and reducing the output data size to 14% of the input raw data size when notifying events.*

## 1. INTRODUCTION

In the last few decades, we have seen many advances in computing technologies, both in hardware miniaturization, data communication, and software solutions, enabling a scenario for using "smart" devices embedded in the most diverse areas of daily life. Nowadays, many healthcare, energy grids, cities, transportation, agriculture, and industry domains use connected sensors, devices, and machines autonomously communicating via the Internet [1, 2, 3, 4, 5]. Each domain area has its particularities and constraints, demanding different resources while sensing, processing, transmitting and presenting data [6, 7].

While IoT is an environment where smart devices such as gadgets and home appliances are interconnected or communicate with cloud-hosted services, Industrial IoT (IIoT) lays over a particular scenario. The industrial environment differs from the other in factors such as the high number of sensors and the need for short data processing response time, besides a harsh environment [8]. In manufacturing, erroneous data and outliers may appear due to sensor noise, communication errors, process disturbances, instrument degradation, mechanical faults, human-related errors, and so on [8]. When an application processes this corrupted sensor data, the overall performance of the system is compromised, making it inaccurate and unreliable. Taking wrong decisions in a manufacturing process can cause out-of-specification products, machinery damage, and even work accidents with human victims [9].

For detecting sensor faults and outliers, we can use Statistical, Nearest-Neighbor, Artificial Neural Network, Cluster-Based, Classification-Based techniques, and so on [10, 11]. Most of the existing methods to process sensor data rely on cloud architecture and Stream Processing (SP) or Complex Event Processing (CEP) services, bringing some problems to industrial plants [12, 13]. Sometimes, companies operating in remote places such as the countryside, offshore or underground do not have reliable and stable Internet access [14]. Usually, IIoT applications execute real-time analysis in product supply chain management, performance evaluation, and simulation [1]. In these cases, data processing is performed by heterogeneous IoT devices on the network edge, with limited processing, memory, and communication capabilities [15].

In Fog computing, data processing and analysis are performed by gateway devices at the network edge, reducing bandwidth requirements, latency, and the need for communicating data to external servers [16]. Deploying fog nodes directly within the network fabric pushes processing even further to the network edge, bringing the fog computing layer closer to the smart end-devices such as sensors and actuators [17]. This approach decreases latency and increases the autonomy of the subsystems since the calculation and decisions are performed locally, and depend mainly on the device's perception of the situation.

Analysing the literature, we found challenges addressed to IoT, such as inaccurate data, lacking Internet access, and real-time applications [14]. However, a significant challenge is the development of infrastructure containing a common framework. Most proposed frameworks cannot be reused for all types of data since they were designed specifically for a particular domain [1]. The studies presented in this paper approached anomaly detection in the network edge applied to a specific area, including underground mining [14], automotive assembly plant [18], water quality monitoring [19], and industrial machinery monitoring [8]. Authors used diverse techniques, such as K-means and C-means [14, 18], Confidence Interval and Interval Variance [20], FFT over vibration and ANN [18], One-Class Classifier SVM, Isolation Forest and Elliptic Envelope [19], Principal Component Analysis (PCA) and R-PCA [21], Chi-Square Distance [8] and Hierarchical Temporal Memory [22].

As identified in the literature, the lacking of standardization in IoT application development, the heterogeneity of IoT hardware and data formats, the variety and complexity in implementing data analytic functions in the fog are the motivations of the present work. To address this challenges, we present STEAM++, a framework for real-time data stream processing and decision-making in the network edge, targeting hardware-limited devices. Although it is very simple to develop a program using the STEAM++ framework, it allows the design of rich solutions regarding data collection and analysis, event detection, and publication of results for external applications and services. Figure 1 represents a high-level overview of our architecture. On the left side (a), we can see the standard cloud-dependent architecture usually adopted in IoT applications. On the right side (b), we can see the STEAM++ architecture for comparison purposes and have a better understanding of our contribution.

A typical IoT application begins with data production, represented as generic raw data sources transmitted over sensor networks. After collected, raw data are processed by a gateway at the network edge, which usually only encapsulates the data frames in a standard protocol and transmits to client applications using Intranet or Internet. Since we propose to bring data analytics techniques to the network edge applying the fog computing concept, we highlight the Analysis, Enrichment, and Evaluation processes executed on far-edge devices by a STEAM++ application.

Lastly, the client applications are responsible for data consumption and business rules processing, and can be hosted either on LAN or cloud.

Besides the STEAM++ framework, we propose a micro-benchmark methodology for assessing embedded IoT applications, monitoring CPU and memory usage, measuring processing time, and calculating output/input data size ratio. To prove the concepts and feasibility of the STEAM++ model and framework, we implemented two applications for processing real scenarios from a semiconductor industry. We performed an entire data flow, from values sensing, processing and analysing data, detecting relevant events, and finally, publishing results to a dashboard. Using our proposed micro-benchmark, the STEAM++ application running on a Raspberry Pi 3 Model B+ consumed on average less than 500kb RAM and 1.0% of CPU usage, processing up to 239 data packets per second and reduced the output data size to 14% of the input raw data size. The results were encouraging, enabling the development of lightweight, fast, interconnected, and valuable IoT applications built with simple programming commands.

Thus, the contributions of this article are twofold:

- The STEAM++ programming framework, simplifying the development of end-to-end IoT applications for real-time data analytics and decision-making in the edge;
- A micro-benchmark methodology for assessing IoT applications embedded in hardware-limited devices.

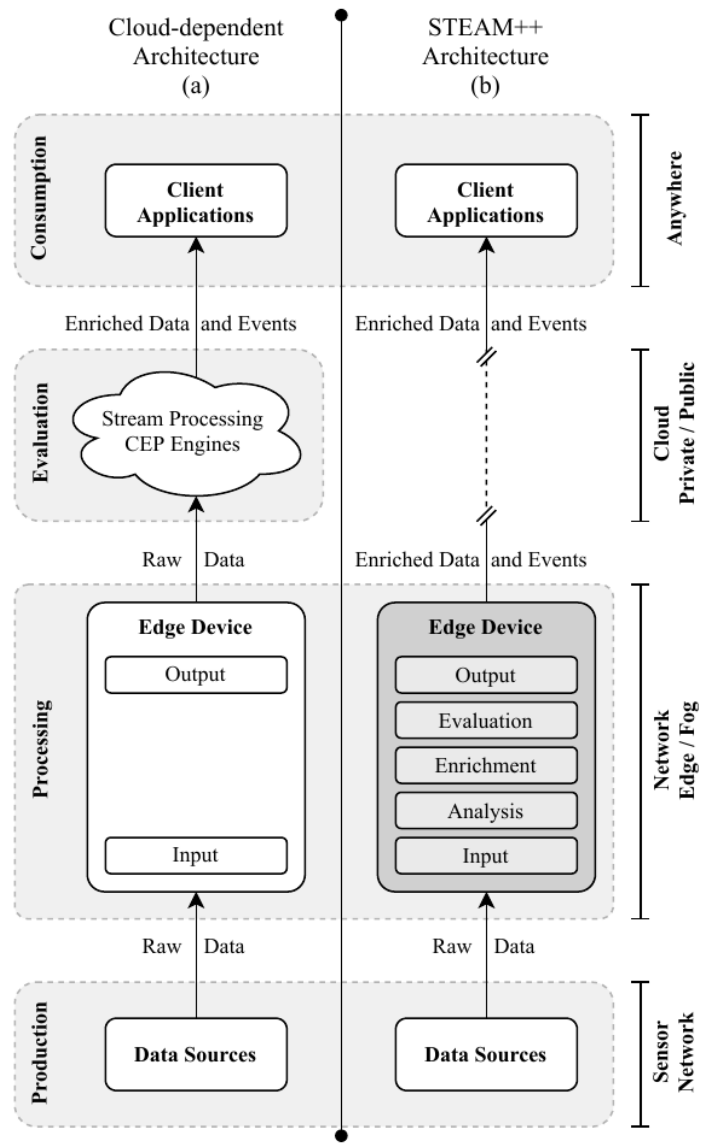

Figure 1. Overview of a Standard Cloud-Dependent IoT Architecture (a)
compared to the STEAM++ Architecture (b)

The rest of the paper is structured as follows. Section 2 is a summary of research and work related to data processing in the edge focusing industrial environment, our proposed model, framework implementation and API are discussed in Section 3, followed by the detailing of the evaluation methodology and experimental study in Section 4. The results are presented and discussed in Section 5, and Section 6 concludes the article.

## 2. RELATED WORK

We analysed the literature aiming to find initiatives performing data analytics in real-time in the network edge. We focused on industrial applications due to their specific conditions regarding harsh environments deployment, susceptible to a variety of interference, noise, and in many cases, without Internet access. The outcome is listed in Table 1. In the sequence, we present a discussion of the main aspects of related works.

Table 1: Related work and their main features

| Reference | Year | Proposal | Anomaly Detection | Noise Removal | Outlier Detection | Real-time Analysis | Techniques |
| --- | --- | --- | --- | --- | --- | --- | --- |
| Liu et al. [14] | 2021 | Algorithm | ✓ | | | | K-Means, C-Means |
| Yin et al. [20] | 2020 | Algorithm | ✓ | | | | Confidence Interval |
| De Vita et al. [18] | 2020 | Architecture | ✓ | | | | FFT, ANN, K-Means |
| Bourelly et al. [19] | 2020 | Algorithm | ✓ | | | | SVM, Isolation Forest, Elliptic Envelope |
| YR et al. [21] | 2020 | Framework | ✓ | | ✓ | ✓ | PCA, R-PCA |
| Liu et al. [8] | 2020 | Algorithm | ✓ | ✓ | | | Chi-Square Distance |
| Greco et al. [22] | 2019 | Architecture | ✓ | | | ✓ | HTM, Node-RED, Flink, Kafka |

Liu et al. [14] proposed an anomaly detection method using K-Means and C-Means over a sliding window, executed on a sink node on the network edge. They monitored multiple sensors in real-time inside an underground mine. In [20], Yin et al. developed an algorithm for anomaly detection using confidence interval, interval variance, and median of a sliding window over a sensor data set. This algorithm computed on the network edge also could distinguish the source of the abnormality. Aiming anomaly detection in an automotive assembly plant, De Vita et al. [18] developed an architectural framework using FFT over vibration, ANN, and K-Means techniques. Bourelly et al. proposed an algorithm for anomaly detection in water quality monitoring [19]. They used One-Class Classifier SVM, Isolation Forest, and Elliptic Envelope for detecting a predefined set of substances commonly considered as dangerous and indicative of an anomalous use of water.

In [21], YR and Champa developed a framework for data aggregation and outlier detection, processing data from 54 sensors, claiming that sensors' inaccuracies and noise make it difficult to define and anticipate data behaviour. They used Principal Component Analysis (PCA) and R-PCA. Liu et al. [8] presented an algorithm computing chi-square distance over a sliding window performing anomaly detection and noise removal for Industrial IoT sensor data in a manufacturing process. Sensors installed in the compressor collected data on temperature, speed, and vibration. For processing wearable sensor data streams, Greco et al. [22] developed an edge-stream computing infrastructure enabling real-time analysis on data coming from wearable sensors. They used the Hierarchical Temporal Memory algorithm, Node-RED, Apache Flink, and Apache Kafka.

## 3. STEAM++ MODEL

In this section, we present STEAM++, a model and framework designed to enable real-time data analytics, decision-making, and data streams enrichment at the network edge. We first presented STEAM in [23], therefore, the current work is an extension of the previous one with three main enhancements. The first improvement of STEAM++ is the Evaluation layer, bringing the decision-making to the Fog and eliminating the cloud dependency. Second, we enhanced the framework's class library, simplifying the development of applications by adding new classes. Last, we propose a micro-benchmark methodology for assessing IoT applications embedded in limited devices on the network edge. The STEAM++ project is hosted on GitHub, and the source code is available at *https://github.com/steam-project/steam*.

## 3.1. Architecture

There are several steps between reading raw data from a sensor until the detection of an event such as anomaly, noise or outlier. Next, we present how STEAM++ performs stream processing in the edge, from data capturing from sensors until providing enriched streams and event detection to client applications. Figure 2 depicts a detailed view of STEAM++ architecture. It consists of a five-layered framework for the development of applications targeting resource-limited devices located at the network edge. Following, we describe each layer in detail.

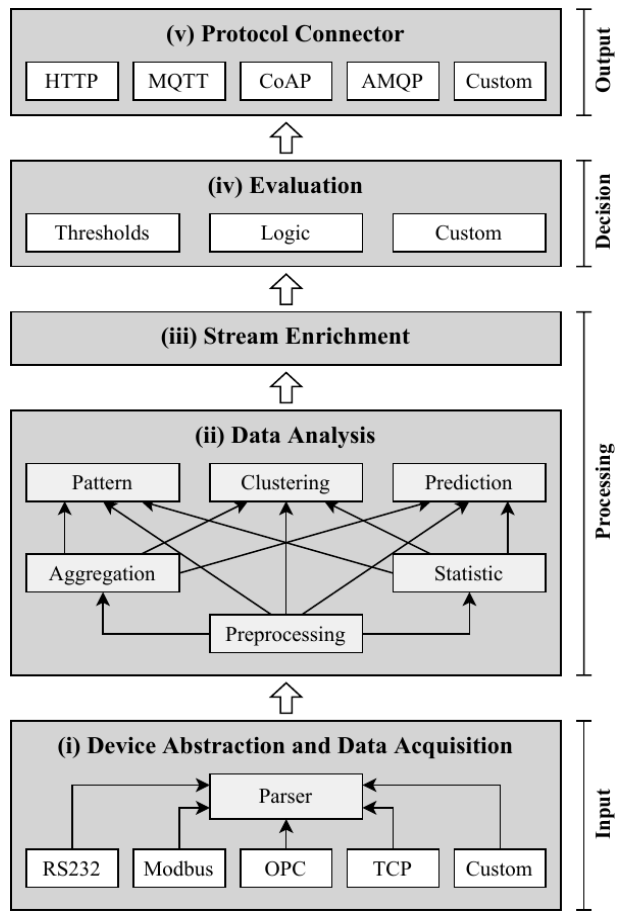

Figure 2. Detailed STEAM++ architecture

- *Device Abstraction and Data Acquisition*: This is the input layer, responsible for capturing data from sensors and far-edge devices in many formats and protocols, parsing, standardizing, and forwarding data streams to the processing step;
- *Data Analysis*: This is the processing step, a layer that provides a set of data analysis techniques, such as filtering, transformation, pattern recognition, outlier detection, prediction, etc. We can also develop custom data analysis functions;
- *Stream Enrichment*: This layer is intended to merge the outcome of the previously mentioned *Data Analysis* layer along with the original data streams, generating an enriched data packet;

International Journal of Computer Science & Information Technology (IJCSIT) Vol 14, No 1, February 2022

- *Evaluation*: The fourth layer evaluates rules, logic, threshold comparing, and performs custom analysis to provide event detection and decision-making. For instance, in this step, we can identify behaviours, noise, outliers, and decide whether or not to send alert messages to client applications or commands to actuators located in the sensor network in specific situations;
- *Protocol Connector*: The output layer is the *Protocol Connector*, responsible for providing output data streams in a standard format and using different communication protocols, enabling client applications to access data in a standard and transparent manner. In this step, a STEAM++ application can publish data sending messages directly to client applications, integration services, message brokers, dashboards, actuators, etc.

### 3.2. Micro-benchmark Methodology and Metrics

For the assessment of the STEAM++ applications, we are proposing a micro-benchmark methodology and three metrics: *CPU/Memory Usage*, *Processing Time* and *Output/Input Ratio*, depicted in Figure 3.

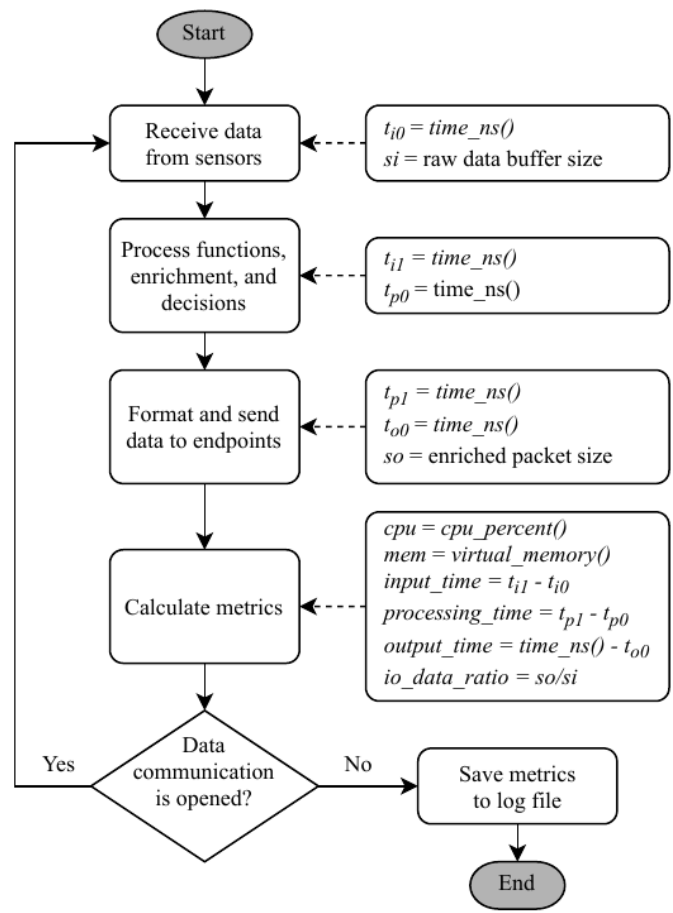

Figure 3: Micro-benchmark methodology and metrics for assessing the STEAM++ framework

To read the system's CPU and memory usage, we are using *cpu_percent()* and *virtual_memory()* methods from *psutil* Python's library, respectively. These values are measured at the end of the processing flow for each received packet, indicating the CPU consumption and memory usage during the tasks. For the *Processing Time* metric, we are measuring the time spent per each STEAM++'s processing layer per packet, from reading raw data until the dispatching of the enriched packet. For this metric, we are using *time_ns()* method from *time* Python's library, that returns an integer number of nanoseconds since the epoch. We are also measuring the total amount of bytes received from sensors, and after, sent to external applications. With these information, we calculate the *Output/Input Data Size Ratio*, indicating the increasing or decreasing factor over the data stream size obtained as the result of STEAM++ processing. When the application ends, the micro-benchmark saves a log file containing the collected and calculated metrics for each processed data packet, where each line represents one data packet processed by the STEAM++ application and the columns represent the metrics, separated by *tab* characters.

### 3.3. Framework Classes

The STEAM++ framework was developed in Python 3.8 as a class library, depicted in Figure 4. Following, we present each class and its functionality.

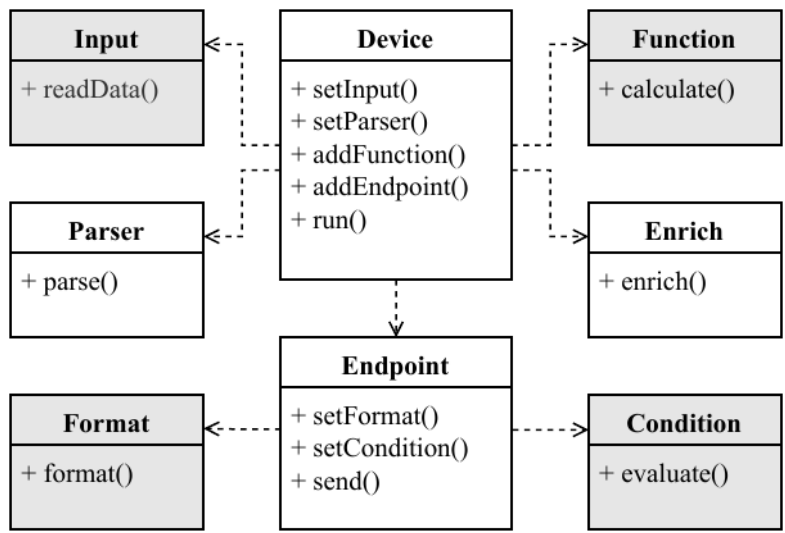

Figure 4: STEAM++ framework class diagram. The main class is *Device*. The classes highlighted in gray color are new, indicating enhancements in the framework comparing with the previous version.

- **Device**: The main class of a STEAM++ application, storing the data, processing logical and analytical functions, and organizing the entire execution flow;
- **Input**: Abstract class that interfaces with the sensor network. The STEAM++ framework extends this class to create specific data acquisition features, supporting several industrial communication protocols such as RS232, Modbus, OPC, etc;
- **Parser**: The default *Parser* class handles raw data frames with a single or multiple values, separated by a single character or a string. We can extend this class, adding the capability of interpreting complex raw data frames structures;
- **Function**: Base class for performing data analysis. The STEAM++ framework extends this class to provide a rich class library. Until this moment, we implemented the following classes: *Min*, *Max*, *Sum*, *Count*, *Mean*, *Median*, *EWMA*, *StDev*, *Slope*, *Arima*, and *Equation;*
- **Enrich**: Class that handles the data stream enrichment process, updating the raw data packets from sensors with the processed data returned from analytical functions;
- **Condition**: Class that evaluates a condition, indicating the occurrence of an event. The STEAM++ framework provides the *EquationCondition*, *MissingValueCondition* and *ThresholdCondition* classes. We can extend the *Condition* class to provide customized evaluation conditions and complex event detection;

- **Format**: Class that formats the enriched data packet before sending it to client applications. The STEAM++ framework extends this class providing a set of formats, such as *MessageFormat*, *CSVFormat*, *TSVFormat*, *JSONFormat*, and *WSO2Format;*
- **Endpoint**: Base class for implementing the output layer of a STEAM++ application, defining the destination of the processed and enriched data streams, messages and events. The STEAM++ framework provides the *FileEndpoint* and *HTTPEndpoint* classes, enabling file storage and HTTP post capabilities, respectively. We can extend this class to create custom publication services, implementing protocols such as MQTT, AMQP, and CoAP, for instance.

## 4. EXPERIMENTAL STUDY

In order to assess the STEAM++ framework, we developed two applications for monitoring the dew-point temperature in a microchip manufacturer's production chain. The infrastructure used in the experiments and the applications are described below.

### 4.1. Infrastructure

The infrastructure used in the experiments is depicted in Figure 5. In this scenario, the sensor network consists of one node (*SN-Node*) receiving measurements from 3 sensors (*Sensors*) at a transmission rate of 1 measurement per second per sensor. The *SN-Node* relays the data to the STEAM++ application running in a Raspberry Pi 3 Model B+ 1GB RAM with Raspbian OS (*IoT Device*) through a raw TCP connection. The TCP data frame consists of an ASCII string containing 6 fields separated by *tab* characters, as follow:

- **id**: Sequential identification of measurement. Integer;
- **timestamp**: Timestamp of measurement. ISO-8601;
- **unit**: Dew-point temperature measurement unit. String;
- **s1**: Dew-point temperature of sensor 1. Float;
- **s2**: Dew-point temperature of sensor 2. Float;
- **s3**: Dew-point temperature of sensor 3. Float;

The STEAM++ applications receive, process, and publish the data, both saving a local log file and sending it to a Node-RED dashboard running in a laptop (*Terminals*) connected to the local network via Wi-Fi. Both dashboard applications simply receive data through an HTTP service and display it in a line chart or text area, without performing any data processing.

### 4.2. Applications

Writing STEAM++ applications is very simple compared to build a from-the-scratch IoT application. The STEAM++ framework provides a set of ready-to-use classes and built-in functions, making it unnecessary to use structured code, complex logic, loops, and conditional statements. The classes' relationships ensure the consistency of the application, assigning to the developer only the task of configuring parameters and objects binding. Figure 6 illustrates the basic application used in the experiments. Line 2 is the *Device* object instantiation, configured to manage a sliding window with the last 20 sensor's measurements. Line 5 defines the data *Input* method as a TCP communication on port 5000. Lines 8 to 11 create the *Parser* object, setting a *tab* character as a values separator and identifying the columns' names. Lines 14 to 16 configure the *HTTPEndpoint*, that consists of the Node-RED chart's input stream URL, format the data output as *JSONFormat*, finally binding the objects to the *Device*. Line 19 starts the application execution.

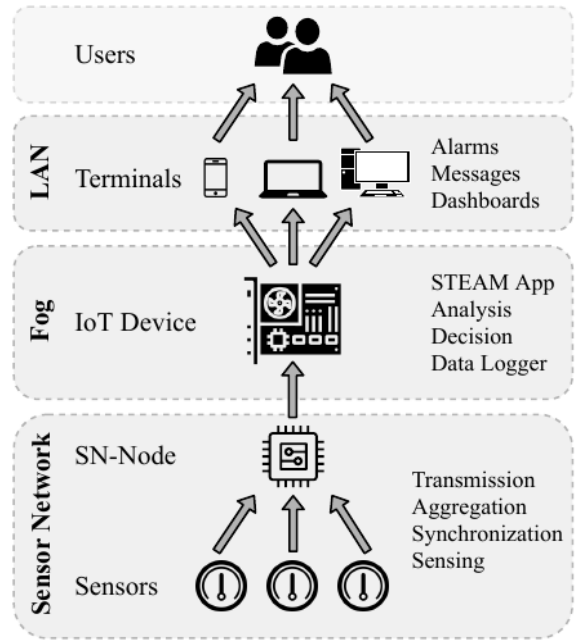

Figure 5: Infrastructure for evaluating the STEAM++ applications

```
1   # Create the Device object
2   device = Device(batchlen=20)
3
4   # Define the Input method
5   device.setInput(TCPInput(port=5000))
6
7   # Create data parser
8   device.setParser(
9     Parser(
10      separator='\t',
11      columns=['id', 'timestamp', 'unit', 's1', 's2', 's3']))
12
13  # Create the endpoint - Node-RED - Line chart - JSON format
14  ep_data = HTTPEndpoint(url='http://node-red-server:1880/datastream')
15  ep_data.setFormat(JSONFormat())
16  device.addEndpoint(ep_data)
17
18  # Run the application
19  device.run()
```

Figure 6: Basic application developed with the STEAM++ framework

Starting from the previous base code, we designed two applications to assess the STEAM++ framework. Both receive data from a sensor node, perform statistical functions, evaluate expressions, and finally enrich the data stream with the outcome of the processed data. Then, the applications send the enriched data stream to a Node-RED dashboard, plotting line charts and displaying relevant event messages. Following, we explain the two applications in detail.

#### 4.2.1. One sensor

This application, whose source code is depicted in Figure 7, receives one single measurement value from one sensor per second and initially computes the standard deviation (line 35) and moving average (line 34) over a sliding window of the last 20 values, corresponding to 20

International Journal of Computer Science & Information Technology (IJCSIT) Vol 14, No 1, February 2022

seconds of measurements. For detecting anomalies in the data stream, we are using a *Statistical Process Control* (SPC) technique, based that common sources of variations result in a normal distribution of samples, where the mean *m* and standard deviation *σ* can be estimated, configured on lines 37 to 41. Any observation outside this control range computed by *m± 3σ* is considered abnormal [24], and reported as a warning message, coded from line 2 to 19. All these values and messages are stored in a local log file (lines 22 to 31) and sent to a remote Node-RED dashboard that plots a line chart and displays the warning messages.

```
1   # Node-RED - Above dynamic upper threshold - Message format
2   ep_upper = HTTPEndpoint(url='http://node-red-server:1880/msgstream')
3   con_upper = ThresholdCondition(columns='value', upper='upper')
4   fmt_upper = MessageFormat(
5       '''<font color=red>{timestamp} -
6       Value {value:.2f} above {upper:.2f}</font><br>''')
7   ep_upper.setCondition(con_upper)
8   ep_upper.setFormat(fmt_upper)
9   device.addEndpoint(ep_upper)
10
11  # Node-RED - Below dynamic lower threshold - Message format
12  ep_lower = HTTPEndpoint(url='http://node-red-server:1880/msgstream')
13  con_lower = ThresholdCondition(columns='value', lower='lower')
14  fmt_lower = MessageFormat(
15      '''<font color=blue>{timestamp} -
16      Value {value:.2f} below {lower:.2f}</font><br>''')
17  ep_lower.setCondition(con_lower)
18  ep_lower.setFormat(fmt_lower)
19  device.addEndpoint(ep_lower)
20
21  # Save to log file - Above dynamic upper threshold - Message format
22  ep_upper_file = FileEndpoint('log_above_upper.txt')
23  ep_upper_file.setCondition(con_upper)
24  ep_upper_file.setFormat(fmt_upper)
25  device.addEndpoint(ep_upper_file)
26
27  # Save to log file - Below dynamic lower threshold - Message format
28  ep_lower_file = FileEndpoint('log_below_lower.txt')
29  ep_lower_file.setCondition(con_lower)
30  ep_lower_file.setFormat(fmt_lower)
31  device.addEndpoint(ep_lower_file)
32
33  # Analytical function list
34  device.addFunction(fn.Mean(format='{:.2f}'))
35  device.addFunction(fn.StDev(format='{:.2f}'))
36
37  device.addFunction(
38      fn.Equation(id='upper', format='{:.2f}', equation='mean + 3 * stdev'))
39
40  device.addFunction(
41      fn.Equation(id='lower', format='{:.2f}', equation='mean - 3 * stdev'))
```

Figure 7: *One Sensor* source code application developed with the STEAM++ framework

#### 4.2.2. Multiple sensors

The multiple sensors application, whose source code is depicted in Figure 8, reads the input data stream containing the measurements from three sensors and detects missing values (line 3). After, from line 21 to 26 it calculates the instantaneous dew-point temperature slope, comparing the current value against the previous measurement for each sensor. Since the sensors monitor the same industrial process, the disagreement of slew rate among the measurements indicates an anomaly, defined between lines 29 and 33. A Node-RED dashboard hosted on a laptop connected to the factory's administrative network receives the values captured from sensors besides the data computed by the STEAM++ application. A line chart plots the measurements of each sensor, and

41

International Journal of Computer Science & Information Technology (IJCSIT) Vol 14, No 1, February 2022

a text area displays event messages such as missing measurements, out-of-threshold values, and slope disagreements, configured on lines 2 to 8 and 11 to 17.

```python
 1  # Node-RED - Missing value - Message format
 2  ep_missing = HTTPEndpoint(url='http://node-red-server:1880/msgstream')
 3  con_missing = MissingValueCondition(columns=['s1', 's2', 's3'])
 4  fmt_missing = MessageFormat(
 5      '<font color=blue>{timestamp} - Value missing</font><br>')
 6  ep_missing.setCondition(con_missing)
 7  ep_missing.setFormat(fmt_missing)
 8  device.addEndpoint(ep_missing)
 9
10  # Node-RED - Slope disagreement - Message format
11  ep_slope = HTTPEndpoint(url='http://node-red-server:1880/msgstream')
12  con_slope = EquationCondition('slope_disagree')
13  fmt_slope = MessageFormat(
14      '<font color=red>{timestamp} - Slope disagreement</font><br>')
15  ep_slope.setCondition(con_slope)
16  ep_slope.setFormat(fmt_slope)
17  device.addEndpoint(ep_slope)
18
19  # Analytical function list
20  # Slope of each sensor measurements
21  device.addFunction(
22      fn.Slope(id='s1_slope', batchlen=2, attribute='s1', format='{:.1f}'))
23  device.addFunction(
24      fn.Slope(id='s2_slope', batchlen=2, attribute='s2', format='{:.1f}'))
25  device.addFunction(
26      fn.Slope(id='s3_slope', batchlen=2, attribute='s3', format='{:.1f}'))
27
28  # Slope disagreement detection - Equation
29  device.addFunction(
30      fn.Equation(
31          id='slope_disagree',
32          equation='''max(s1_slope, s2_slope, s3_slope) > 0.1 and
33                      min(s1_slope, s2_slope, s3_slope) < -0.1'''))
```

Figure 8: *Multiple Sensors* source code application developed with the STEAM++ framework

## 5. RESULTS AND DISCUSSION

This section presents the detailed findings from the experiments, making a profile of the STEAM++ applications' behaviour in the specified scenarios.

### 5.1. Dashboards

The first result is the user's point-of-view, in other words, two Node-RED dashboards for data visualization containing a line chart and a display of relevant events. Figure 9 depicts the *One Sensor Application*, described in subsection 4.2.1. The upper and lower lines are the dynamic upper and lower thresholds respectively, computed by $m \pm 3\sigma$ equation. The centralized green line is the moving average $m$, and the oscillating blue line is the dew-point temperature read from the sensor. The chart also shows a red circle where the dew-point temperature exceeds the upper threshold, and a blue circle where the temperature gets below the lower threshold. On the right side of the chart, a text area displays messages containing the events detected by the application. In blue are the warnings related to low values, and in red are the messages associated with high values.

42

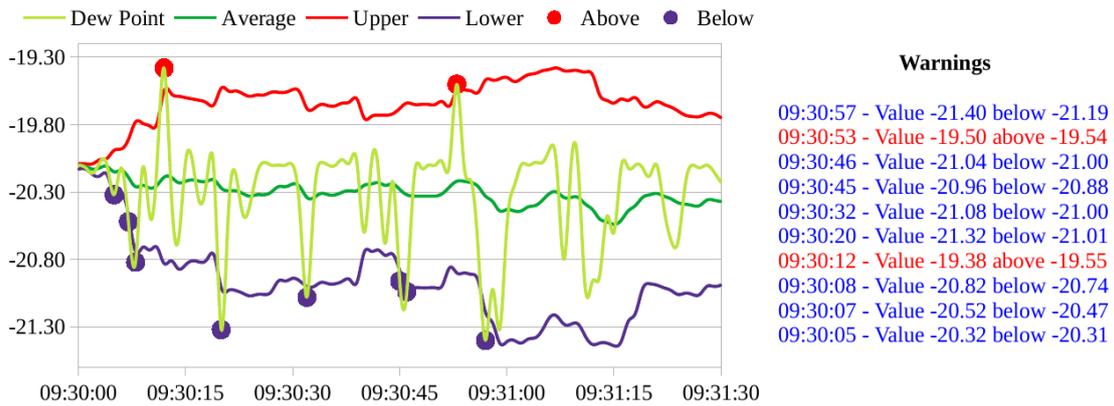

Figure 9: Node-RED dashboard screenshot for one sensor application

The dashboard of *Multiple Sensors Application* is illustrated in Figure 10. The chart is plotting three lines, representing three sensors of dew-point temperature. The blue circles are pointing missing values, indicating the absence of a sensor reading or a transmission failure, causing the lacking of values in the time series. The vertical red lines are indicating slope disagreement among the sensor's measurements, as detailed in subsection 4.2.2. As the previous dashboard, this one also displays warning messages. In blue are the value missing alerts and in red color are the slope disagreement messages.

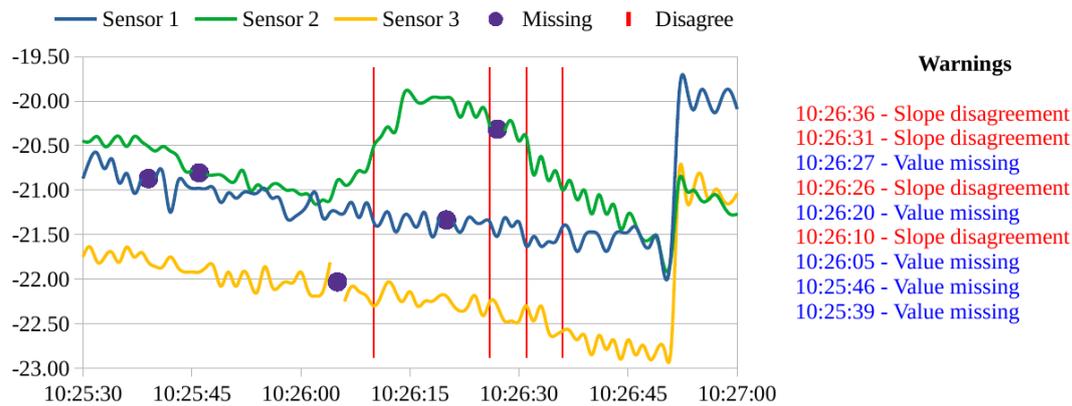

Figure 10: Node-RED dashboard screenshot for multiple sensors application

## 5.2. CPU and Memory Usage

Due to limited computational resources, CPU and memory usage are key indicators in the IoT environment. To have a significant and reliable overview of resource consumption, we executed each application 30 times, collecting the instant system's CPU load and the overall used memory. Figures 11 and 12 depicts a typical *One Sensor* and *Multiple Sensors* application behaviours respectively, regarding CPU and memory consumption. In both scenarios, the average CPU load is below 1% with peaks less than 2.5%, and the average memory usage is less than 500kb, with peaks below 800kb, excluding the outliers. The exact values are detailed in Table 2 and the data distribution is presented in Figure 13.

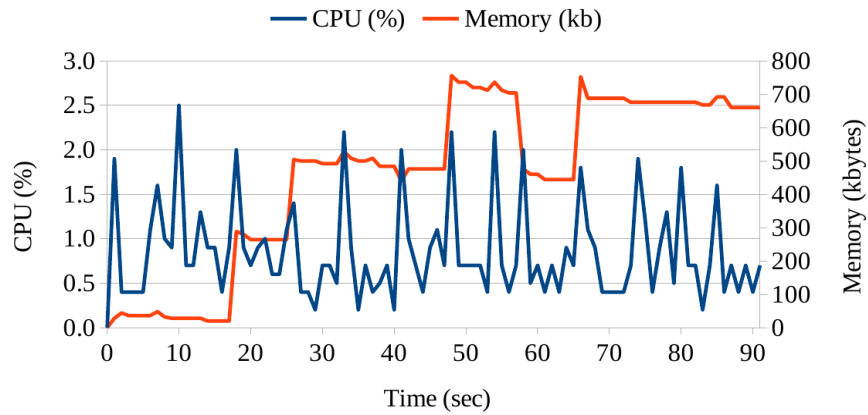

Figure 11: Typical CPU and Memory usage behaviour for a test with one sensor

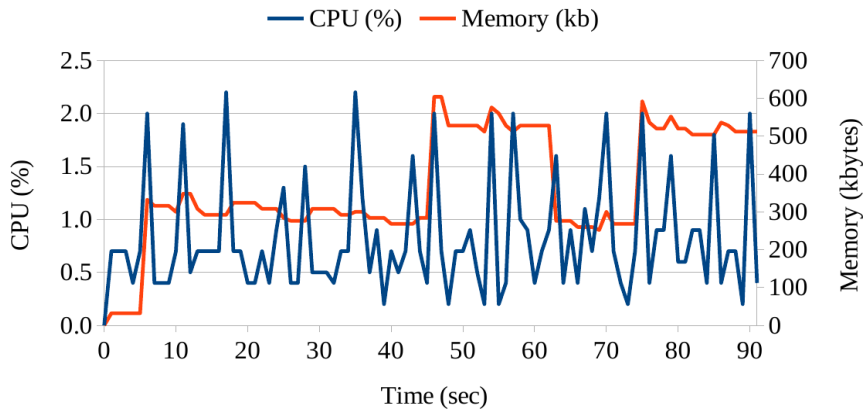

Figure 12: CPU and Memory usage for one test of multiple sensors application

Table 2: Average CPU and Memory usage - compilation of 30 experiments with one packet per second

| Application | One Sensor | | Multiple Sensors | |
|---|---|---|---|---|
| Metric | CPU (%) | Memory (kb) | CPU (%) | Memory (kb) |
| Minimum | 0.804 | 193.41 | 0.685 | 168.25 |
| Maximum | 0.955 | 1127.93 | 0.814 | 851.64 |
| Average | 0.867 | 523.82 | 0.741 | 435.64 |
| Median | 0.864 | 496.92 | 0.744 | 395.78 |
| 1st quartile | 0.846 | 377.49 | 0.724 | 291.05 |
| 3rd quartile | 0.886 | 590.72 | 0.757 | 581.72 |

## 5.3. Processing Time

The experiments initially performed in this work for assessing time used a processing rate of 1 packet per second. Applying the *Processing Time* metric described in subsection 3.2, we collected the time spent in *Input*, *Processing* and *Output* layers. Figure 14 presents the distribution of time spent per processing layer. The *Input* step, responsible for collecting and parsing the raw data from sensors, is the fastest of all, consuming 728μs on average. The

International Journal of Computer Science & Information Technology (IJCSIT) Vol 14, No 1, February 2022

*Processing* layer, that performs calculations and evaluates conditions, used on average 5554μs to complete the tasks. *Output*, the slower layer, consumed on average 108997μs to format and send data to the endpoints, which in this case, consisted of saving a local log file and sending to Node-RED dashboard via HTTP. Proportionally, the *Input* process took 0.63%, the *Processing* layer consumed 4.82%, and the *Output* registered 94.55% of the time spent for processing the packets. Table 3 presents the detailing of processing time metric, and Figure 15 depicts the data distribution.

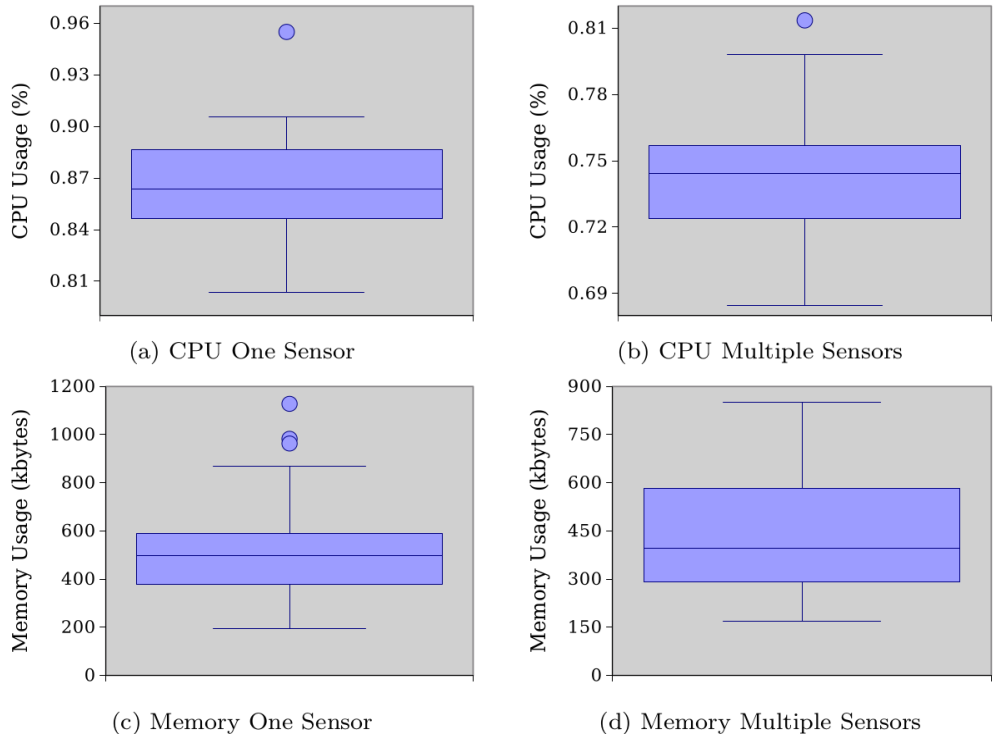

Figure 13: CPU and Memory distribution - compilation of 30 experiments with one packet per second

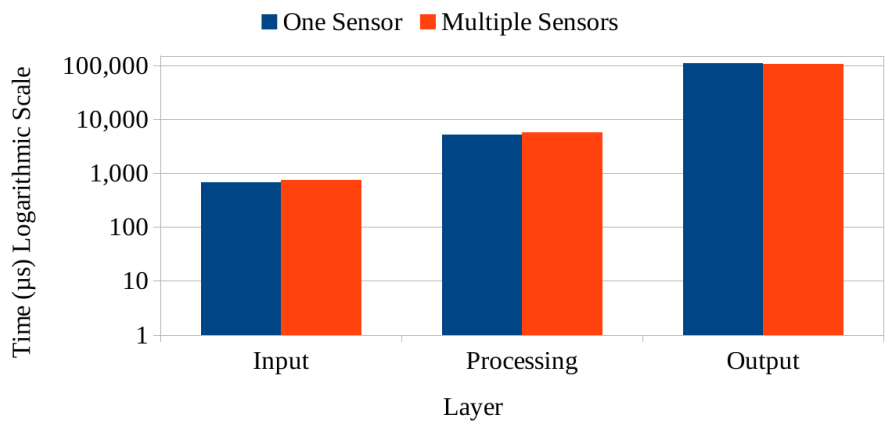

Figure 14: Average time spent per processing layer - 30 experiments with one packet per second - logarithmic scale

Table 3: Average processing time per layer - 30 experiments with one packet per second

| Application | One Sensor | | | | Multiple Sensors | | | |
|---|---|---|---|---|---|---|---|---|
| Metric (μs) | Input | Processing | Output | Total | Input | Processing | Output | Total |
| Minimum | 674 | 5143 | 99760 | 105801 | 746 | 5727 | 98169 | 104713 |
| Maximum | 698 | 5431 | 117701 | 123527 | 810 | 6120 | 115166 | 121722 |
| Average | 687 | 5239 | 109886 | 115811 | 768 | 5870 | 108069 | 114707 |
| Median | 688 | 5235 | 110025 | 115995 | 763 | 5850 | 108880 | 115446 |
| 1st quartile | 680 | 5199 | 108107 | 114061 | 758 | 5823 | 106168 | 112775 |
| 3rd quartile | 693 | 5275 | 112295 | 118200 | 773 | 5884 | 109873 | 116548 |

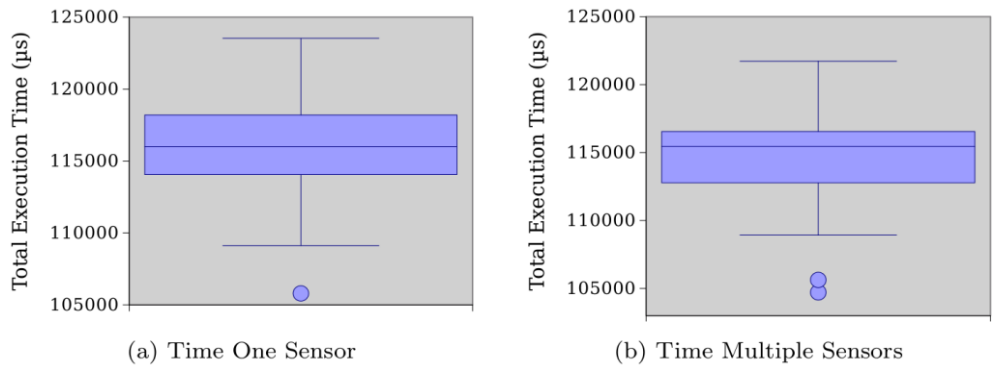

(a) Time One Sensor

(b) Time Multiple Sensors

Figure 15: Total time distribution - compilation of 30 experiments with one packet per second

## 5.4. Output / Input Ratio

The raw data received from sensors usually present a small size, containing only relevant information in a simple structure. The processing flow calculates new values that are included in the data packet and formatted as a JSON structure, containing symbols, delimiters, and identifiers. Moreover, data sent to client applications can assume diverse structures and formats such as XML, HTML, and CSV, increasing the size of the published data. If on the one hand the processed data packet has increased in size, on the other hand the STEAM++ application can evaluate conditions and send only relevant messages to the client applications. This feature acts like a filter, and can drastically reduce the amount of transmitted data, and consequently, decrease the network traffic.

The data processing flow depicted in Figure 16 demonstrate the differences in formats and sizes comparing one single data input and its corresponding output data packet for Multiple Sensors experiments. The STEAM++ application receives the raw data packet, performs calculations and assembles the enriched data packet in JSON format. Next, the application sends the packet to the Node-RED dashboard, and at the same time, converts it to a *Tab Separated Values* string (TSV) and saves it in a log file.

In Table 4, we present the differences in data sizes detailed by output method and application. Compared with the raw data, the TSV formatted log file size increases between 141.73% and 182.22% due to the inclusion of the calculated values. However, the publishing to the chart dashboard requires a JSON format, resulting in the increment from 528.65% to 608.18% in whole data size, compared to raw data. Nevertheless, we only send messages to the dashboard's text area when we detect an event. It acts as a filter over the processed data, decreasing the whole size of published data. In this case, the overall output data stream decreased to 14.23% and 18.65% sizes compared to the input raw data stream.

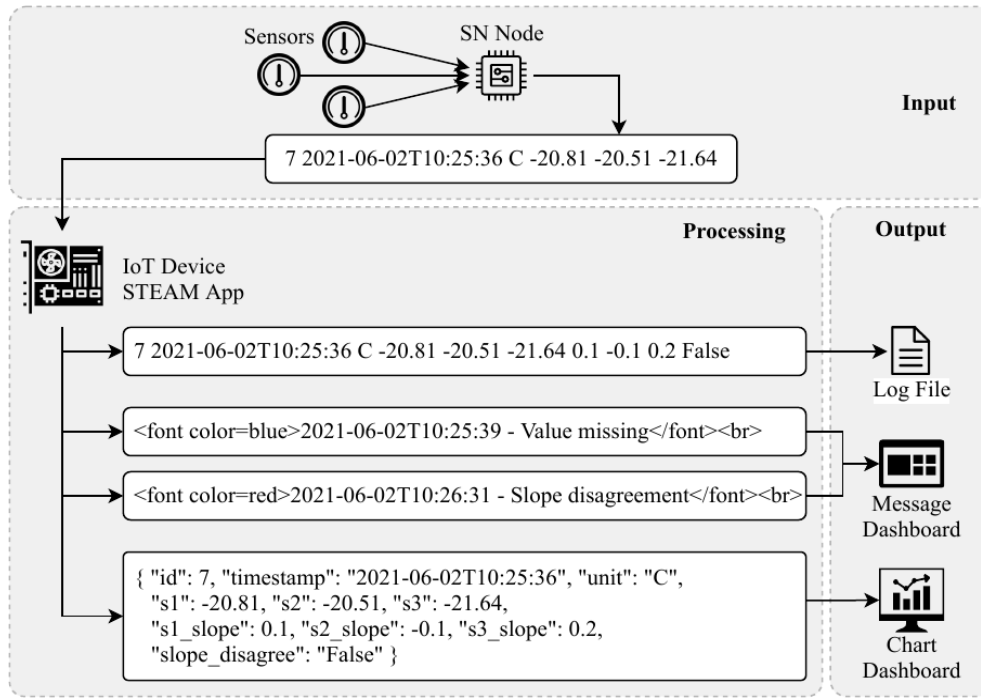

Figure 16: Data transformation during the processing flow - *Multiple Sensors*

Table 4: Output / Input data size ratio

| Application | One Sensor | | Multiple Sensors | |
| --- | --- | --- | --- | --- |
| Metric | Size (bytes) | Ratio | Size (bytes) | Ratio |
| Raw Data Input | 2874 | - | 4146 | - |
| Log File | 5237 | 182.22% | 5876 | 141.73% |
| Dashboard Chart | 17479 | 608.18% | 21918 | 528.65% |
| Dashboard Message | 536 | 18.65% | 590 | 14.23% |

## 5.5. Testing to the limits

Until this moment, the applications implemented and executed in the experiments presented a low consumption of CPU and memory, running a processing rate of 1 data packet per second. However, one expected contribution of the STEAM++ framework is enabling the development of near real-time IoT applications. To identify the limits of the processing speed and computational resource consumption, we stored the sensor's data in a text file. Then, we fed the application at the highest data flow rate it could process. We repeated this test 30 times to obtain a reliable result set.

In the first stress test scenario, we used the same *Multiple Sensors Application* detailed in Subsection 4.2, but we simulated the sensors reading and forced the data transmission to the limit of the speed. Figure 17 illustrates one typical test, which the average CPU load reached 15.4% with peaks of 33.3%, and the average memory consumption was 527.04kb. Considering all the 30 tests, the CPU load registered 15.4% and memory 289.81kb in average.

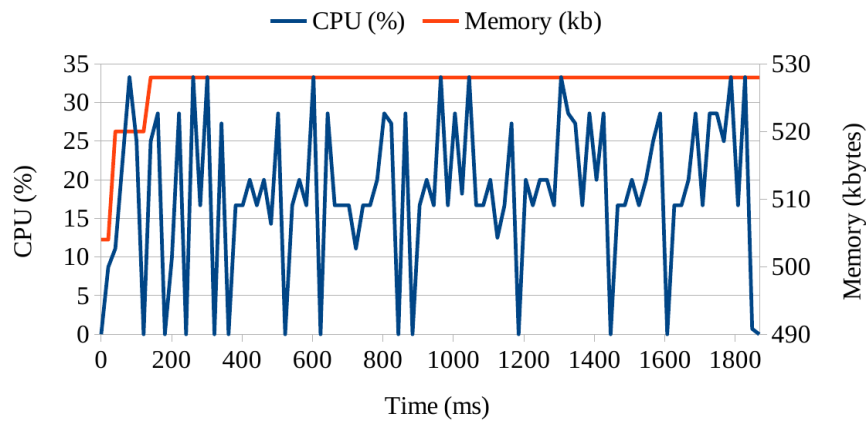

Figure 17: CPU and Memory usage for a high speed data flow - Publishing data and messages to a dashboard

For the second stress test, we removed the publication of data and messages to the Node-RED dashboard, since the HTTP network communication is a typically slow task comparing to accessing local resources. In this scenario, we only saved the processed data and messages to a local log file. We depicted an arbitrary test case in Figure 18, but we also performed the test 30 times. The average CPU load for this specific test case reached 22.4% with peaks of 66.7%, and the average memory usage was 271.74kb. Compiling all the 30 tests, the CPU load reached 18.0% and memory consumption was 196.91kb in average.

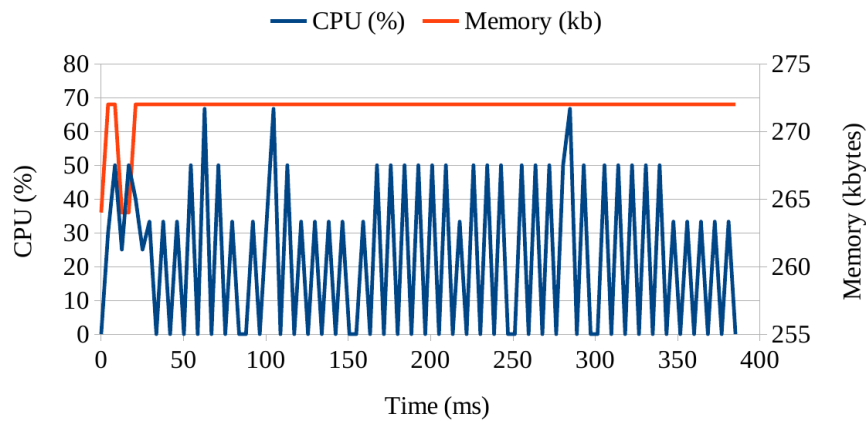

Figure 18: CPU and Memory usage for a high speed data flow - Saving data and messages to a local log file

Regarding processing time, we completed the first test in 1827.578 ms, and the second one in 380.895 ms on average. We identified that this time difference refers to publishing data to the dashboard hosted on the remote computer. Converting these measurements to packets processed per time, the first scenario could handle 49.79 packets per second, while the second reached the rate of 238.91 packets per second. In other words, when sending data to the dashboard, each packet consumed 20083μs, and when saving to a local log file, the same process lasted 4186μs. Table 5 presents the compilation of time spent per each processing layer collected from the 30 stress tests.

Table 5: Average processing time per layer - 30 stress test experiments

| Output | Dashboard | | | | File | | | |
|---|---|---|---|---|---|---|---|---|
| Metric (μs) | Input | Processing | Output | Total | Input | Processing | Output | Total |
| Minimum | 345 | 3133 | 11995 | 15503 | 354 | 3306 | 317 | 3993 |
| Maximum | 427 | 3891 | 52688 | 57005 | 409 | 3553 | 421 | 4298 |
| Average | 369 | 3302 | 16413 | 20083 | 372 | 3467 | 347 | 4186 |
| Median | 366 | 3275 | 12800 | 16448 | 372 | 3497 | 339 | 4214 |
| 1st quartile | 360 | 3217 | 12629 | 16234 | 366 | 3430 | 332 | 4120 |
| 3rd quartile | 369 | 3315 | 13724 | 17460 | 377 | 3518 | 360 | 4251 |

Comparing the stress test with the real-case test, more specifically the output step, we noticed a significant disagreement between the times elapsed on sending data to the Node-RED dashboard. In the real case test, while the average time taken by the output step was 108978μs, this same step performed in the stress test consumed 16413μs on average, processing exactly the same data. Analysing the network status with the *netstat* command, we identified inactive connections between the Raspberry Pi and the Node-RED while sending one packet per second, illustrated in Figure 19. However, we witnessed three established and no inactive connections on the stress tests, depicted in Figure 20. The need for establishing new connections after the one-second wait results in an overall time increasing measured on the output layer, however, it does not happen in the stress test that uses the same connections over the entire test.

```
Proto    Foreign Address        State           PID/Program name
tcp      192.168.1.120:55550    ESTABLISHED     13174/node-red
tcp      192.168.1.120:55556    TIME_WAIT
tcp      192.168.1.120:55558    TIME_WAIT
tcp      192.168.1.120:55560    TIME_WAIT
tcp      192.168.1.120:55562    TIME_WAIT
tcp      192.168.1.120:55564    FIN_WAIT2
tcp      192.168.1.120:55566    TIME_WAIT
tcp      192.168.1.120:55568    FIN_WAIT2
```

Figure 19: Network status of a real-case test

```
Proto    Foreign Address        State           PID/Program name
tcp      192.168.1.120:55570    ESTABLISHED     13174/node-red
tcp      192.168.1.120:55572    ESTABLISHED     13174/node-red
tcp      192.168.1.120:55574    ESTABLISHED     13174/node-red
```

Figure 20: Network status of a stress test

## 6. CONCLUSIONS

Aiming the particularities of the Industrial IoT, this article presented STEAM++, a framework to simplify the development of end-to-end IoT applications for real-time data analytics and decision-making in the edge, besides the capability of publishing processed data and events to a variety of services and formats. We also proposed a micro-benchmark methodology for assessing embedded IoT applications, monitoring CPU and memory usage, measuring processing time, and calculating output/input data size ratio.

One remarkable aspect of writing STEAM++ applications is its simplicity compared to build a from-the-scratch IoT application. The framework provides a set of ready-to-use classes and built-in functions, making it unnecessary to use structured code, complex logic, loops, and conditional statements. This feature enables even non-programmers the possibility to develop rich IoT applications by simply configuring parameters.

To show the feasibility of the STEAM++ framework, we implemented two real-case applications in a semiconductor industry and achieved consistent outcomes. Since one of our goals was to build lightweight fog computing solutions, we obtained on average less than 1.0% of CPU load and less than 436kb of memory consumption, besides fast response times, processing up to 239 data packets per second, reducing the output data size to 14% of the input raw data size when notifying events, and integrating with a remote dashboard application.

The IoT is spreading daily and evolving to diverse areas such as healthcare, transportation, agriculture, and industry, facing specific situations and challenging requirements. To reinforce the fog computing principles, in future research, we intend to extend the STEAM++ framework to other IoT areas, propose a scalable architecture to deal with a dynamic data processing demand, and develop more analytic and communication modules, consolidating all data processing in the network edge.